\documentclass[twocolumn,secnumarabic,amssymb, amsmath, nofootinbib,tightenlines,
nobibnotes, aps, prl,epsfig]{revtex4}
\usepackage{graphicx}
\usepackage{dcolumn}
\usepackage{bm}
\begin{document}
\preprint{APS/123-QED}
\title{Evolution of the Longitudinal Structure Function at Small-$x$}

\author{G.R.Boroun }
\altaffiliation{boroun@razi.ac.ir; grboroun@gmail.com}
\affiliation{ Physics Department, Razi University, Kermanshah
67149, Iran}
\date{\today}

\begin{abstract}
We derive an approximation approach to evolution of the
longitudinal structure function, by using a Laplace-transform
method. We solve the master equation and derive the longitudinal
structure function as a function of  the initial condition
$F_{L}(x,Q_{0}^{2})$ at small-$x$. Our results are independent of
the longitudinal coefficient functions and extend from the
leading-order (LO) up to next-to-next-to-leading order (NNLO). The
comparisons with H1 data and other parameterizations are made and
results show that they are in agreement with H1 data and some
phenomenological models.
\end{abstract}
 \pacs{11.55Jy, 12.38.-t, 14.70.Dj}
\keywords{Longitudinal structure function; Gluon distribution;
QCD; Small-$x$; Regge- like behavior} 
\maketitle
\subsection{1 Introduction}
The measurement of the longitudinal structure function
$F_{L}(x,Q^{2})$ is of great theoretical importance, since it may
allow us to distinguish between different models describing the
QCD evolution at small-$x$. In deep- inelastic scattering (DIS),
the structure function measurements remain incomplete until the
longitudinal structure function $F_{L}$ is actually measured [1].
The longitudinal structure function in deep inelastic scattering
is one of the observable from which the gluon distribution can be
unfolded.\\
 At small-$x$ values, the dominant contribution to
$F_{L}(x,Q^{2})$ comes from the gluon operators. Hence a
measurement of $F_{L}(x,Q^{2})$ can be used to extract the gluon
structure function and therefore the measurement of $F_{L}$
provides a sensitive test of perturbative QCD [2,3]. At this
region, the longitudinal structure function can be related to the
gluon and sea- quark distribution.
 In principle, the data on the singlet part
 of the structure function $F_{2}$ constrain the sea quarks
  and the data on the slope $\frac{dF_{2}}{d{\ln}Q^{2}}$
  determine the gluon density. Moreover, the longitudinal
  structure function $F_{L}(x,Q^{2})$ can be related at small-$x$
  with structure function $F_{2}$ and the derivation
  $dF_{2}/d{\ln}Q^{2}$[4-9]. In this way most precise predictions
  based on data of $F_{2}$ and
  $dF_{2}/d{\ln}Q^{2}$ can be obtained for $F_{L}$. These
  predictions can be considered as indirect experimental data for
  $F_{L}$.\\
The behavior of the structure function $F_{2}$ at small $x$ and
large $Q^{2}$ have been discussed considerably over the past years
[10] into the double asymptotic scaling. Double asymptotic scaling
follows from a computation [11-12] of the asymptotic form of the
structure function $F^{p}_{2}(x,Q^{2})$ at small $x$ based on the
use of the operator product expansion and renormalization group at
leading perturbative order. It thus relies only on the assumption
that any increase in $F^{p}_{2}(x,Q^{2})$ at small $x$ is
generated by perturbative QCD evolution, rather than being due to
some other (nonperturbative) mechanism manifested by an increase
in the starting distribution $F^{p}_{2}(x,Q_{0}^{2})$. A better
understanding of QCD physics at small x have been achieved by
considering more observables and thus (over-)constraining the
definite of the parton densities [13]. In Ref.13 the interplay
between perturbative and non-perturbative dynamics has been
pointed out in the context of QCD analysis of the small $x$
behavior of the proton structure function $F^{p}_{2}(x,Q^{2})$.
The factorization theorem of mass singularities provides a
representation of $F_{2}$ in terms of phenomenological parton
densities and perturbatively computable splitting and coefficient
functions. Then, S.Catani have been considered the physical
anomalous dimensions relating the singlet components of $F_{2}$
and $F_{L}$. That solve it for Mellin transformations of the
parton densities. In N-space, an available program that deal with
DGLAP evolution is QCD PEGASUS [14], which is  a parton
distribution functions (PDFs) evolution program based on
Mellin-space inversion.\\
To study the longitudinal structure function we use  the
Laplace-transform technique for solving the Altarelli- Martinelli
equation[15]. In recent years, Laplace-transform technique have
proved to be valuable tools for the solving of the DGLAP [16-18]
evolution equations in the LO approximation up to NLO [19-23].
Here, a similar procedure is used to derive evolution of the
longitudinal structure function inside the proton. We obtain an
analytical solution for evolution of the longitudinal structure
function at small $x$ in terms of the initial condition at the
starting scale $Q_{0}^{2}$. Thus we can determine the longitudinal
structure function at small-$x$ directly as a function of the
initial longitudinal structure function, and this result is
independent of the knowledge about the coefficient functions for
$F_{L}$ at LO up
to NNLO. \\
 The content of our paper is as follows. In  section $2$ we
 describe the basic theory to extract the longitudinal structure function from the gluon distribution function
  at small-$x$. Section $3$  is devoted to the analytical solution of the master equation  for the longitudinal structure
  function by Laplace-transform technique.
Finally, an analytical analysis of our solution is presented  and
the obtained results are compared with other methods which are
followed by
 results and discussions.\\

\subsection{2. Basic Theory }
We specifically consider the longitudinal structure function
$F_{L}$, projected from the hadronic tensor by combination of the
metric and the spacelike momentum transferred by the virtual
photon $(g_{\mu\nu}-q_{\mu}q_{\nu}/q^{2})$. As it is proportional
to hadronic tensor as follows
\begin{equation}
F_{L}(x,Q^{2})/x=\frac{8x^{2}}{Q^{2}}p_{\mu}p_{\nu}W_{\mu\nu}(x,Q^{2}),
\end{equation}
where $p^{\mu}(p^{\nu})$ is the hadron momentum and $W^{\mu\nu}$
is the hadronic tensor. In this relation we neglecting the hadron
mass.\\
The basic hypothesis is that the total cross section of a hadronic
process can be written as the sum of the contributions of each
parton type (quarks, antiquarks, and gluons) carrying a fraction
of the hadronic total momentum. In the case of deep- inelastic-
scattering it reads
\begin{equation}
d\sigma_{H}(p)=\sum_{i}{\int}dyd\hat{\sigma}_{i}(yp)\Pi_{i}^{0}(y),
\end{equation}
where $d\hat{\sigma}_{i}$ is the cross section corresponding to
the parton $i$ and $\Pi_{i}^{0}(y)$ is the probability of finding
this parton in the hadron target with the momentum fraction $y$.
Now, taking into account the kinematical constrains one gets the
relation between the hadronic and the partonic structure functions
\begin{eqnarray}
f_{j}(x,Q^{2})&=&\sum_{i}{\int}_{x}^{1}\frac{dy}{y}\textsf{f}_{j}(\frac{x}{y},Q^{2})\Pi_{i}^{0}(y)\\\nonumber
&&=\sum_{i}\textsf{f}_{j}{\otimes}\Pi_{i}^{0}(y)\hspace{0.5cm},j=2,L,
\end{eqnarray}
where $\textsf{f}_{j}(x,Q^{2})=F_{j}(x,Q^{2})/x$ and the symbol
${\otimes}$ denotes convolution according to the usual
prescription,
$f(x){\otimes}g(x)=\int_{x}^{1}\frac{dy}{y}f(y)g(\frac{x}{y})$.
Equation (3) expresses the hadronic structure functions as the
convolution of the partonic structure function, which are
calculable in perturbation theory, and the probability of finding
a parton in the hadron which is a nonperturbative function. At
small values of $x$, $F_{L}$ is driven mainly  by gluons through
the transition $g{\rightarrow}{\hspace{0.1cm}}q\overline{q}$
($g(x,Q^{2})$ is the gluon density). Recently [24], the process
$gg {\rightarrow}gg$ where the external gluons are on-shell have
been obtained by calculation of the forward jet vertex at
next-to-leading order in the BFKL formalism, offering an explicit
derivation of the gluon-initiated contribution. As this adds to
that previously [25] calculated for the quark-initiated vertex and
completes the derivation of the full vertex.\\
 Therefore $F_{L}$
can be used for the extraction of the gluon distribution in the
proton, and it provides a crucial test of the validity of
perturbative QCD in this kinematical range. So, in correspondence
with Eq.(3) one can write Eq.(1) for the gluon density dominated
at small-$x$ values by follows
\begin{eqnarray}
F^{g}_{L}/x{=}\frac{\alpha_{s}}{4\pi}[\textsf{f}_{L,G}^{(LO+....)}{\otimes}g^{0}],
\end{eqnarray}
where $\textsf{f}_{L,G}^{,}$s  are the LO up to  NNLO partonic
longitudinal structure function corresponding to gluons [26-27].
We present the expressions, after full agreement has been
achieved, in the form of kernels $C_{L,G}$ which give $F_{L}$ upon
convolution with
 the gluon distribution
\begin{eqnarray}
F^{g}_{L}(x,Q^{2})=\frac{\alpha_{s}(Q^{2})}{4\pi}<e^{2}>C_{L,G}(\alpha_{s},\frac{x}{y}){\otimes}G(y,Q^{2})\\\nonumber
=\frac{\alpha_{s}(Q^{2})}{4\pi}<e^{2}>\int_{x}^{1}\frac{dy}{y}C_{L,G}(\alpha_{s},\frac{x}{y})G(y,Q^{2}),
\end{eqnarray}
where $C_{L,G}(x,Q^{2})$ is the DIS coefficient function for
$F^{g}_{L}(x,Q^{2})$. The average squared charge (=5/18 for even
$n_{f}$) is represented by $<e^{2}>$, where $n_{f}$ denotes the
number of effectively massles flavours. We write the perturbative
expansion of the coefficient functions as
\begin{eqnarray}
C_{L,G}(\alpha_{s},x)&=&c_{L,G}^{\rm
LO}(x)+\frac{\alpha_{s}(Q^{2})}{4\pi}c_{L,G}^{\rm
NLO}(x)\nonumber\\
&& +(\frac{\alpha_{s}(Q^{2})}{4\pi})^{2} c_{L,G}^{\rm NNLO}(x).
\end{eqnarray}
The longitudinal coefficient functions are given in Refs.[28-30].
The running coupling constant $\frac{\alpha_{s}}{4\pi}$ has the
form in the LO, NLO and NNLO respectively [31]
\begin{equation}
\frac{\alpha_{s}^{\rm LO}}{4\pi}=\frac{1}{\beta_{0}t},
\end{equation}
\begin{equation}
\frac{\alpha_{s}^{\rm
NLO}}{4\pi}=\frac{1}{\beta_{0}t}[1-\frac{\beta_{1}{\ln}t}{\beta_{0}^{2}t}],
\end{equation}
and
\begin{eqnarray}
\frac{\alpha_{s}^{\rm
NNLO}}{4\pi}&=&\frac{1}{\beta_{0}t}[1-\frac{\beta_{1}{\ln}t}{\beta_{0}^{2}t}+\frac{1}{(\beta_{0}t)^{2}}
[(\frac{\beta_{1}}{\beta_{0}})^{2}\nonumber\\
&&(\ln^{2}t-{\ln}t+1)+\frac{\beta_{2}}{\beta_{0}}]].
\end{eqnarray}
where $\beta_{0}=\frac{1}{3}(33-2n_{f})$,
$\beta_{1}=102-\frac{38}{3}n_{f}$ and
$\beta_{2}=\frac{2857}{6}-\frac{6673}{18}n_{f}+\frac{325}{54}n_{f}^{2}$
are the one-loop,two-loop and three-loop corrections to the QCD
$\beta$-function. The variable $t$ is defined as
$t={\ln}(\frac{Q^{2}}{\Lambda^{2}})$ and $\Lambda$ is the QCD
cut- off parameter.\\
Therefore the master equation for the longitudinal structure
function can be written as
\begin{eqnarray}
(\frac{\alpha_{s}}{4\pi}<e^{2}>)^{-1}F^{g}_{L}(x,Q^{2}){\equiv}\mathcal{F}^{g}_{L}(x,Q^{2})\\
=\int_{x}^{1}\frac{dy}{y}C_{L,G}(\alpha_{s},\frac{x}{y})G(y,Q^{2}).\nonumber
\end{eqnarray}
This equation for the longitudinal structure function has the
explicit dependence to the gluon distribution function. To extract
an approximation solution of the longitudinal structure function
evolution from the DGLAP equation without dependence to the gluon
distribution, we should solve Eq.10 for the evolution of the
longitudinal structure function $F_{L}(x,Q^{2})$ into
$F_{L}(x,Q^{2}_{0})$. Let us consider the differential form of
Eq.5 as for simplicity we assume that the running coupling
constant in $C_{L,G}(\alpha_{s},x)$ is constant at NLO up to NNLO.
Therefore we have
\begin{eqnarray}
\frac{{\partial}F^{g}_{L}}{{\partial}{\ln}Q^{2}}=\frac{<e^{2}>}{4\pi}\frac{d\alpha_{s}}{d{\ln}Q^{2}}C_{L,G}(\frac{x}{y}){\otimes}G(y,Q^{2})\\\nonumber
+\frac{\alpha_{s}}{4\pi}<e^{2}>C_{L,G}(\frac{x}{y}){\otimes}\frac{{\partial}G(y,Q^{2})}{{\partial}{\ln}Q^{2}},
\end{eqnarray}
where the derivative of the gluon distribution  with respect to
$\ln Q^{2} $ i.e. ${\partial}G(x,Q^{2})/{\partial}\ln Q^{2}$,  at
small-$x$ is given by the DGLAP evolution equation as we have
[15-18,32-35]
\begin{eqnarray}
\frac{{\partial}G(x,Q^{2})}{{\partial}{\ln}Q^{2}}=\frac{\alpha_{s}}{4\pi}{\int_{x}^{1}}\frac{dy}{y}[
P_{gg}(\frac{x}{y},\alpha_{s}(Q^{2})) G(y,Q^{2})],
\end{eqnarray}
where the splitting functions $P_{ij}^{,}s$ are the LO, NLO and
NNLO Altarelli- Parisi splitting kernels as
\begin{eqnarray}
P_{gg}(x,\alpha_{s}(Q^{2}))=P_{gg}^{\rm
LO}(x)+\frac{\alpha_{s}(Q^{2})}{4\pi}P_{gg}^{\rm
NLO}(x)\nonumber\\
+(\frac{\alpha_{s}(Q^{2})}{4\pi})^{2} P_{gg}^{\rm NNLO}(x).
\end{eqnarray}
\subsection{3. Master equation for evolution of the longitudinal structure function at small-$x$}
To evolution of the longitudinal structure function, we follow the
procedure that was used by authors Refs.[19-22] and employ the
Laplace-transform method to solve Eqs.10-12. Now we use the
coordinate transformation as
\begin{eqnarray}
\upsilon &{\equiv}& \ln(1/x).
\end{eqnarray}
In $\upsilon$-space, Eq.10 appears as
\begin{eqnarray}
\mathcal{\widehat{F}}^{g}_{L}(\upsilon,Q^{2})=\int_{0}^{\upsilon}\widehat{C}_{L,G}(\upsilon-w)\widehat{G}(w,Q^{2})dw,
\end{eqnarray}
where the functions $\mathcal{\widehat{F}}^{g}_{L}$,
$\widehat{C}_{L,G}$ and $\widehat{G}$ are given by
\begin{eqnarray}
\mathcal{\widehat{F}}^{g}_{L}(\upsilon,Q^{2}){\equiv}\mathcal{{F}}^{g}_{L}(e^{-\upsilon},Q^{2}),\nonumber\\
\widehat{C}_{L,G}(\upsilon,Q^{2}){\equiv}{C}_{L,G}(e^{-\upsilon},Q^{2}),\nonumber\\
\widehat{G}(\upsilon,Q^{2}){\equiv}{G}(e^{-\upsilon},Q^{2}).
\end{eqnarray}
If we take the Laplace-transform of Eq.15, then we have
\begin{eqnarray}
{\mathcal{L}}[\mathcal{\widehat{F}}^{g}_{L}(\upsilon,Q^{2});s]~~~~~~~~~~~~~~~~~~~~~~~~~~~~~~~~~~~~~~~~~~~\nonumber\\
={\mathcal{L}}[\int_{0}^{\upsilon}\widehat{C}_{L,G}(\upsilon-w)\widehat{G}(w,Q^{2})dw;s].
\end{eqnarray}
Therefore
\begin{eqnarray}
{{F}}^{g}_{L}(s,Q^{2})=h(s){g}(s,Q^{2}).
\end{eqnarray}
Also, derivative of the longitudinal structure function (Eq.11) in
$\upsilon$-space appears as
\begin{eqnarray}
\frac{{\partial}{{F}}^{g}_{L}(s,Q^{2})}{{\partial}{\ln}Q^{2}}=\frac{d{\ln}\alpha_{s}}{d{\ln}Q^{2}}{{F}}^{g}_{L}(s,Q^{2})
+\frac{\alpha_{s}}{4\pi}{{F}}^{g}_{L}(s,Q^{2}){\Phi_{g}}(s).
\end{eqnarray}
At small-$x$ derivative of the gluon distribution  function
(Eq.12) in $\upsilon$-space is straightforward [22], as we can be
written this equation  by this form
\begin{eqnarray}
\frac{{\partial}g(s,Q^{2})}{{\partial}{\ln}Q^{2}}{=}\frac{\alpha_{s}}{4\pi}\Phi_{g}(s)g(s,Q^{2}),
\end{eqnarray}
where $g(s)=
{\mathcal{L}}[\hat{G}(\upsilon);s]=\int_{0}^{\infty}\hat{G}(\upsilon)e^{-s\upsilon}d\upsilon
$ ($\hat{G}(\upsilon){\equiv}G(e^{-\upsilon})$). The coefficient
function $\Phi_{g}(s)$ at LO is given by [22]
\begin{eqnarray}
\Phi^{LO}_{g}(s)=\frac{33-2n_{f}}{3}+12(\frac{1}{s}-\frac{2}{1+s}-+\frac{1}{2+s}\\\nonumber
-\frac{1}{3+s}-\psi(1+s)-\gamma_{E})
\end{eqnarray}
where $\psi(x)$ is the digamma function and
$\gamma_{E}=0.5772156...$ is Euler$_{^{,}}$s constant.\\
For obtain an approximation form for the evolution of the
longitudinal structure function at small-$x$, we rewrite Eq.19 in
$s$-space as
\begin{eqnarray}
\frac{{\partial}{{\ln
F}}^{g}_{L}(s,Q^{2})}{{\partial}{\ln}Q^{2}}=\frac{d{\ln}\alpha_{s}}{d{\ln}Q^{2}}
+\frac{\alpha_{s}}{4\pi}{\Phi_{g}}(s).
\end{eqnarray}
In the above equation we take the inverse Laplace transform using
the known inverse
${\mathcal{L}}^{-1}[F(s,Q^{2});\upsilon]=\widehat{F}(\upsilon,Q^{2})$,
we find that
\begin{eqnarray}
\frac{{\partial}{{\ln
\widehat{F}}}^{g}_{L}(\upsilon,Q^{2})}{{\partial}{\ln}Q^{2}}=\frac{d{\ln}\alpha_{s}}{d{\ln}Q^{2}}\delta(\upsilon)
+\frac{\alpha_{s}}{4\pi}{\widehat{\Phi}_{g}}(\upsilon),
\end{eqnarray}
where
${\mathcal{L}}^{-1}[{\Phi_{g}}(s);\upsilon]={\widehat{\Phi}_{g}}(\upsilon)={{\Phi}_{G}}(x)$.\\
The solution of the evolution equation of the longitudinal
structure function in terms of the initial values of function
$F_{L}(x,Q_{0}^{2})$ is straightforward. Finally we have
\begin{eqnarray}
F^{g}_{L}(x,Q^{2})=F_{L}(x,Q_{0}^{2})\eta(Q^{2},Q^{2}_{0})e^{\tau(Q^{2},Q^{2}_{0}){\Phi}_{G}(x)},
\end{eqnarray}
where
\begin{eqnarray}
\eta(Q^{2},Q^{2}_{0})=\frac{\alpha_{s}(Q^{2})}{\alpha_{s}(Q^{2}_{0})},
\end{eqnarray}
and
\begin{eqnarray}
\tau(Q^{2},Q^{2}_{0})=\frac{1}{4\pi}\int_{Q^{2}_{0}}^{Q^{2}}\alpha_{s}(Q'^{2})d\ln
Q'^{2}.
\end{eqnarray}
This result is an approximation approach to the evolution equation
for the longitudinal structure function  and gives an analytical
expression for the evolution of $F_{L}$ at leading order (LO). We
emphasize that $F^{g}_{L}(x,Q^{2})$ directly is not dependence to
the gluon distribution function and to the longitudinal
coefficient function. It is dependence to the running coupling
constant and to the gluonic splitting function at LO up to NNLO,
as the explicit form of the gluonic splitting function at LO
($\phi^{LO}_{G}(x)$) is
\begin{eqnarray}
\Phi^{\rm LO}_{G}(x)&=&\frac{33-2n_{f}}{3}+12(1-2x+x^2-x^3-\gamma_{E})\nonumber\\
&&+6x(1+\coth(\frac{1}{2}{\ln}\frac{1}{x})).
\end{eqnarray}
This method can be generalized to NLO up to NNLO. The evolution of
the NLO (up to NNLO) splitting coefficients is straightforward,
but the method can not be completely extended because of the
impossibility of analytically inverting the required Laplace
transform to the NLO (up to NNLO) splitting functions needed in
the DGLAP evolution equation (Eq.12) [22]. At the limit of
small-$x$ the two and three-loop splitting functions read [36-38]
\begin{eqnarray}
\Phi^{\rm
NLO}_{G}(x){\rightarrow}4(\frac{12C_FT_F-46C_AT_F+C_A^2(41-3\pi^{2})}{9})
\end{eqnarray}
and
\begin{eqnarray}
\Phi^{\rm
NNLO}_{G}(x)&{\rightarrow}&(14214.2+n_{f}182.958-n_{f}^{2}2.79835)\nonumber\\
&&-(2675.85+n_{f}157.269)\ln(\frac{1}{x})
\end{eqnarray}
with $C_{A}=N_{c}=3$,
$C_{F}=\frac{N_{c}^{2}-1}{2N_{c}}=\frac{4}{3}$ and
 $T_{f}=\frac{1}{2}n_{f}$.\\
\subsection{4. Results and Discussions}
In this paper, we have obtained an analytical solution for the
evolution of the longitudinal structure function  at small-$x$.
Our solution is model independent of the gluon distribution
function and it is free of any longitudinal coefficient function.
It is only dependence on the running coupling constant and gluonic
splitting function based on the Laplace transform technique at
small-$x$. It can be provided as our results at small-$x$ can
predict be the longitudinal structure function to high-$Q^{2}$
values and suggest that the precise measurement of experimental
values for $F_{L}(x,Q^{2})$ over a wide kinematic range of
small-$x$ and high-$Q^{2}$ can be done (Recently the longitudinal
structure function measured by H1-2013 Collaboration [46-48] at
high-$Q^{2}$ values ). To confirm the method
 and results, the calculated values are compared with the $H1$ data on the longitudinal
 structure function. It is shown that, our results are in agreement with
 experimental $H1$ data for $F_{L}$, if one takes into the total
 errors, and is consistent with a higher order QCD calculations of $F_{L}$
 which essentially show increase as $x$ decreases. We observe that the
 calculations results are consistent with the two pomeron model.
 Thus implying that Regge theory and perturbative evolution may be
 made compatible at small-$x$. The result not only gives striking
 support to the two- pomeron description of small $x$ behavior,
 but also a rather clean test of perturbative QCD itself.\\
We computed the predictions for  the longitudinal structure
function in the kinematic range where it has been measured by $H1$
collaboration [4,6-7,46-48] and compared with DL model [39-42]
based on hard Pomeron exchange and also $k_{T}$ factorization [49]
at small $x$. At small-$x$, the longitudinal structure function
receive large logarithmic corrections coming from resummation of
large powers of $\alpha_{s}{\ln}\frac{1}{x}$, where goes beyond
the standard collinear factorization formalism [50-51] using the
unintegrated gluon density obtained from the
Kwiecinski-Martin-Stasto (KMS) approach [52]. This approach
includes important effects of higher order resummation. Using
$k_{T}$ factorization at the on-shell limit which the transverse
momentum of the gluon $k^{2}$ is much smaller than the virtuality
of the photon, $k^{2}{<<}Q^{2}$ and this is consistent with the
collinear factorization. The $k_{T}$ factorization formula can be
determined the inclusive cross section in dipole representation.
Where, the longitudinal structure function is proportional to the
longitudinal polarized photon-proton cross section, or it is
proportional to the color dipole cross section [49,53]. With
respect to the GBW saturation model [54-55] for the dipole cross
section, the leading twist terms are proportional to the linear
terms in the gluon density. Consequently, the leading twist-2 part
in the dipole picture gives the longitudinal structure function as
derived in the Refs.[53-55].\\
 Our analytical predictions are presented
as functions of $x$ for the $Q^{2}=20, 45$ and $200 ~GeV^{2}$.
Here we obtain the gluonic longitudinal structure function
$F_{L}^{g}$ by evolving up in $Q^{2}$ from $F_{L}(x,Q_{0}^{2})$ at
the input scale, $Q_{0}^{2}=1 GeV^{2}$, obtained in an
approximation analysis to the DGLAP evolution equation with the DL
initial starting function. We defined the coupling via the
$n_{f}=4$ definition of $\Lambda_{QCD}$ for the MRST set of
partons[43-45] as the values of $\Lambda_{QCD}$ at LO up to NNLO is displayed in Table 1.\\
 The results are presented in Figs.1-3 where they are compared with the
 recent $H1$ data [46-48] and with the results obtained with the
help of other standard gluon distribution functions. As can be
seen in all figures, the increase of  our calculations for the
 longitudinal structure functions $F^{g}_{L}(x,Q^{2})$ towards
 small-$x$ are consistent with the NLO QCD calculations, reflecting the
 rise of the gluon momentum distribution in this region. This is
 because the hard-Pomeron exchange defined by DL model is expected
 to hold in the small-$x$ limit. Also we compared the longitudinal structure function  with on-shell limit of the $k_{T}$
 factorization, and with the leading twist-2 term from the dipole picture.
  Comparing our results at NNLO with the QCD predictions from other sets is good and these results are
  consistent with previous observations [5,28,43-45,56]. \\
 In conclusion, we have computed  the longitudinal structure function based on
 the Laplace transforms at low-$x$. These calculations allow us to determine the
 gluonic longitudinal structure function at small-$x$, directly from the
 initial distribution at $Q^{2}=Q^{2}_{0}$ where  $Q^{2}_{0}$ is the starting value for the evolution.
The calculations are consistent with the experimental data for H1
collaboration. We compared our result with the $k_{T}$
factorization scheme in the collinear and the dipole limits.
 As an illustration of this paper, we have used the analytical
 solution to the evolution equation to obtain test of the
 consistency of published longitudinal structure function and predict these results to small-$x$ and high-$Q^{2}$
 values as compared with the recently data from H1 Collab. at $Q^{2}=200 GeV^{2}$.\\

\textbf{References}\\
\hspace{2cm}1. A.Gonzalez-Arroyo, C.Lopez, and F.J.Yndurain, phys.lett.B\textbf{98}, 218(1981).\\
\hspace{2cm}2. A.M.Cooper- Sarkar, G.Inglman, K.R.Long,
R.G.Roberts, and D.H.Saxon , Z.Phys.C\textbf{39}, 281(1988).\\
\hspace{2cm}3.  R.G.Roberts, \textit{The structure of the proton}, (Cambridge University Press 1990)Cambridge.\\
\hspace{2cm}4. S.Aid et.al, $H1$ collab. phys.Lett. {\bf B393}, 452-464 (1997).\\
\hspace{2cm}5. R.S.Thorne, phys.Lett. {\bf B418}, 371(1998); arXiv:hep-ph/0511351(2005).\\
\hspace{2cm}6. C.Adloff et.al, $H{1}$ Collab., Eur.Phys.J.C\textbf{21}, 33(2001).\\
\hspace{2cm}7. N.Gogitidze et.al, $H{1}$ Collab., J.Phys.G\textbf{28}, 751(2002).\\
\hspace{2cm}8. A.V.Kotikov and G.Parente, JHEP \textbf{85},
17(1997).\\
\hspace{2cm}9. A.V.Kotikov and G.Parente, Mod.Phys.Lett.A\textbf{12}, 963(1997).\\
\hspace{2cm}10. R.D.Ball and S.Forte, Phys.Lett.B\textbf{77},
336(1994).\\
\hspace{2cm}11. A. De Rujula, S.L. Glashow, H.D. Politzer, S.B.
Treiman, F. Wilczek and A. Zee, Phys. Rev. D\textbf{10}, 1649
(1974).\\
\hspace{2cm}12. Yu.L. Dokshitzer, Sov. Phys. J.E.T.P.\textbf{46},
641(1977).\\
\hspace{2cm}13. S.Catani, Z.Phys.C\textbf{75}, 665(1997).\\
\hspace{2cm}14. A.Vogt, Comp.Phys.Comm\textbf{170}, 65(2005).\\
\hspace{2cm}15. G.Altarelli and G.Martinelli, Phys.Lett.B\textbf{76}, 89(1978).\\
\hspace{2cm}16. Yu.L.Dokshitzer, Sov.Phys.JETP {\textbf{46}},
641(1977).\\
\hspace{2cm}17. G.Altarelli and G.Parisi, Nucl.Phys.B
\textbf{126}, 298(1977).\\
\hspace{2cm}18. V.N.Gribov and L.N.Lipatov,
Sov.J.Nucl.Phys. \textbf{15}, 438(1972).\\
19. M.M.Block, L.Durand, D.W.McKay, Phys.Rev.D\textbf{77},
094003(2008).\\
20. M.M.Block, L.Durand, D.W.McKay, Phys.Rev.D\textbf{79},
014031(2009).\\
21. M.M.Block, Eur.Phys.J.C\textbf{69},
425(2010).\\
22. M.M.Block, L.Durand, P.Ha and D.W.McKay,
Phys.Rev.D\textbf{83},
054009(2011).\\
23. M.M.Block, L.Durand, P.Ha and D.W.McKay,
Eur.Phys.J.C\textbf{65},
1(2010).\\
24. M.Hentschinski, A.Sabio Vera and C.Salas, Phys.Rev.D\textbf{87}, 076005(2013).\\
25. M. Hentschinski and A. Sabio Vera, Phys. Rev. D \textbf{85},
056006(2012).\\
26.D.I.Kazakov, et.al., Phys.Rev.Lett\textbf{65}, 1535(1990).\\
27.J.L.Miramontes, J.sanchez Guillen and E.Zas, Phys.Rev.D \textbf{35}, 863(1987).\\
28. S.Moch, J.A.M.Vermaseren, A.vogt, Phys.Lett.B \textbf{606},
123(2005).\\
29. A.D.Martin, W.J.Stirling, R.S.Thorne, Phys.Lett.B \textbf{635}, 305(2006).\\
30. A.D.Martin, W.J.Stirling, R.S.Thorne, Phys.Lett.B \textbf{636}, 259(2006).\\
31. B.G. Shaikhatdenov, A.V. Kotikov, V.G. Krivokhizhin, and G.
Parente, Phys.Rev.D\textbf{81}, 034008(2010).\\
32. G.G.Callan and D.Gross, Phys.Lett.B\textbf{22}, 156(1969).\\
33. E.B.Zijlstra and  W.L Van Neerven, Nucl.Phys.B\textbf{383}, 552(1992).\\
34. G.R.Boroun and B.Rezaei, Eur.Phys.J.C\textbf{73}, 2412(2013).\\
35. G.R.Boroun and B.Rezaei, Eur.Phys.J.C\textbf{72}, 2221(2012).\\
36. S.Moch, J.Vermaseren and A.Vogt, Nucl.Phys.B \textbf{688}, 101(2004).\\
37. S.Moch, J.Vermaseren and A.Vogt, Nucl.Phys.B \textbf{691}, 129(2004).\\
38. A.Retey, J.Vermaseren , Nucl.Phys.B \textbf{604}, 281(2001).\\
39. A. Donnachie and P.V.Landshoff, Phys.Lett.B\textbf{533},
277(2002).\\
40. A. Donnachie and P.V.Landshoff, Phys.Lett.B\textbf{550},
160(2002).\\
41.  J.R.Cudell, A.
Donnachie and P.V.Landshoff, Phys.Lett.B\textbf{448}, 281(1999).\\
42. P.V.Landshoff, arXiv:hep-ph/0203084.\\
43. A.D.Martin, R.G.Roberts, W.J.Stirling,R.S.Thorne, Phys.Lett.B \textbf{531}, 216(2002).\\
44. A.D.Martin, R.G.Roberts, W.J.Stirling,R.S.Thorne, Eur.Phys.J.C\textbf{23}, 73(2002).\\
45. A.D.Martin, R.G.Roberts, W.J.Stirling,R.S.Thorne, Phys.Lett.B \textbf{604}, 61(2004).\\
46. F.D. Aaron, et al., [H1 Collaboration],
Eur.Phys.J.C\textbf{71}, 1579(2011).\\
47. F.D. Aaron, et al., [H1 Collaboration], Phys.Lett.B\textbf{665}, 139(2008).\\
48. V. Andreev, et al., [H1 Collaboration], arXiv:1312.4821v1 [hep-ex](2013).\\
49. K.Golec-Biernat and A.M.Stasto, Phys.Rev.D{\textbf80},
014006(2009).\\
50. S.Catani, M.Ciafaloni and F.Hautmann, Phys.Lett.B{\textbf242},
97(1990).\\
51. J.C.Collins and R.K.Ellis, Nucl.Phys.B{\textbf360},
3(1991).\\
52. J.Kwiecinski, A.D.martin and A.M.Stasto,
Phys.Rev.D{\textbf56}, 3991(1997).\\
53. J.Bartles, K.Golec-Biernat and K.Peters,
Eur.Phys.J.C{\textbf17},
121(2000).\\
54. K.Golec-Biernat and M.Wusthoff, Phys.Rev.D{\textbf59},
014017(1999).\\
55.  J.Bartles, K.Golec-Biernat and L.Motyka,
Phys.Rev.D{\textbf81},
054017(2010).\\
56. C.Pisano, arXiv:hep-ph/0810.2215.\\
\begin{table}[h]
\centering \caption{The QCD coupling and corresponding $\Lambda$
parameter for $n_{f}=4$, for LO, NLO and NNLO fits according to
Ref.[43-45].}\label{table:table2}
\begin{minipage}{\linewidth}
\renewcommand{\thefootnote}{\thempfootnote}
\centering
\begin{tabular}{|l|c|c|} \hline\noalign{\smallskip}  & $ \alpha_{s}(M_{Z}^{2})$ &$
\Lambda_{QCD}(MeV)$  \\
\hline\noalign{\smallskip}
LO & 0.130 & 220 \\
NLO & 0.119 & 323 \\
NNLO & 0.1155 & 235 \\
\hline\noalign{\smallskip}
\end{tabular}
\end{minipage}
\end{table}
\begin{figure}
\includegraphics[width=0.5\textwidth]{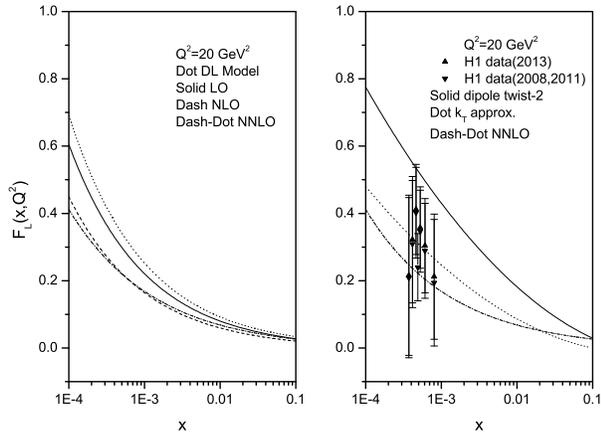}
\caption{Predictions for $F_{L}^{g}(x,Q^{2})$ at $Q^{2}=20
GeV^{2}$ at LO up to NNLO, compared with H1 data [46-48], DL model
[39-42], dipole twist-2 [49,53-55] and $k_{T}$ approximation
[49-52]. }\label{Fig1}
\end{figure}
\begin{figure}
\includegraphics[width=0.5\textwidth]{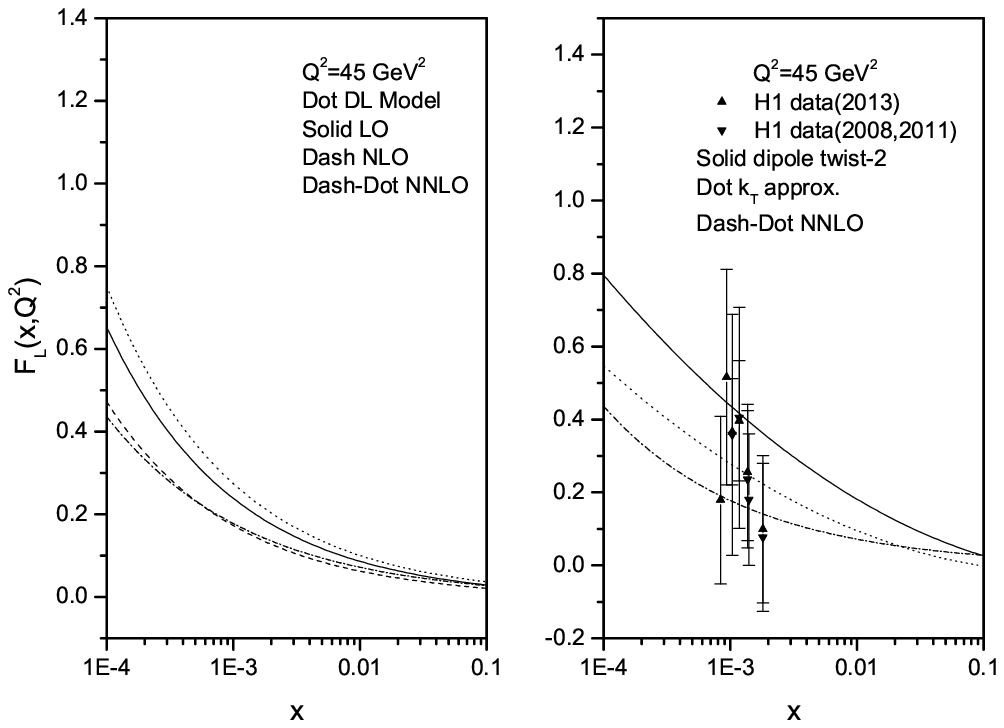}
\caption{The same as Fig.1 at $Q^{2}=45 GeV^{2}$. }\label{Fig2}
\end{figure}
\begin{figure}
\includegraphics[width=0.5\textwidth]{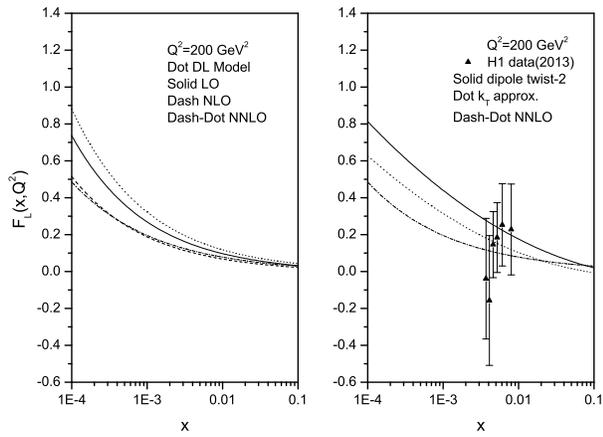}
\caption{The same as Fig.1 at  $Q^{2}=200 GeV^{2}$. }\label{Fig3}
\end{figure}

\end{document}